\documentclass[preprint,amsfonts]{revtex4-1}
\usepackage{amssymb}
\usepackage{amsmath}
\usepackage{color}
\usepackage{graphicx}
\usepackage{natbib}
\usepackage{hyperref}
\definecolor{darkgreen}{rgb}{0.1,0.6,0.3}
\definecolor{darkred}{rgb}{0.6,0.3,0.1}
\hypersetup{
    colorlinks=true,       
    linkcolor=blue,          
    citecolor=darkgreen,        
    filecolor=magenta,      
    urlcolor= black           
}

\begin{document}

\title{Repeatability of evolution on epistatic landscapes}

\author{Benedikt Bauer$^1$}
\email{bauer@evolbio.mpg.de}
\author{Chaitanya S. Gokhale$^2$}
\email{c.gokhale@massey.ac.nz}

\affiliation{%
$^1$Department of Evolutionary Theory, Max Planck Institute  for Evolutionary Biology, August-Thienemann-Stra{\ss}e 2, 24306 Pl\"{o}n, Germany}%

\affiliation{%
$^2$New Zealand Institute for Advanced Study, Massey University, Auckland, New Zealand}%

\begin{abstract}
Evolution is a dynamic process.
The two classical forces of evolution are mutation and selection. 
Assuming small mutation rates, evolution can be predicted based solely on the fitness differences between phenotypes. 
Predicting an evolutionary process under varying mutation rates as well as varying fitness is still an open question. 
Experimental procedures, however, do include these complexities along with fluctuating population sizes and stochastic events such as extinctions.
We investigate the mutational path probabilities of systems having epistatic effects on both fitness and mutation rates using a theoretical and computational framework. 
In contrast to previous models, we do not limit ourselves to the typical \textit{strong selection, weak mutation (SSWM)}-regime or to fixed population sizes. 
Rather we allow epistatic interactions to also affect mutation rates. 
This can lead to qualitatively non-trivial dynamics. 
Pathways, that are negligible in the SSWM-regime, can overcome fitness valleys and become accessible. 
This finding has the potential to extend the traditional predictions based on the SSWM foundation and bring us closer to what is observed in experimental systems.
\end{abstract}


\maketitle

\section*{Introduction}
How repeatable is evolution?
As the metaphor by Stephen J Gould goes `if we run the tape of life back from the start how likely is it that we will get the same outcome that we see around us today?' \cite{beatty:TJP:2006}.
The pioneering work of Lenski et al. tackled this question experimentally with microbes. 
It is now possible to literally play back evolution from a certain starting point and see where it leads \cite{lenski:AmNat:1991,cooper:PNAS:2003,blount:Nature:2012,meyer:Science:2012,travisano:Evolution:2013}.

Such empirical explorations made the until then theoretical concept of fitness landscapes tangible. 
The concept of a fitness landscape is a mapping between the genotype and the phenotype of an organism. 
Since selection acts on the phenotype or essentially on the fitness of the phenotype, the genotype of each phenotype can be attributed a certain fitness. 
Connecting the genotypes which are one mutational step away from each other leads to the concept of fitness landscapes \cite{haldane:PCPS:1927,fisher:book:1930}.
Such empirical studies do make it clear that predictions will not be based on simple rules but complicated phenomena such as epistasis and epigenetics which play a major role in the process of evolution \cite{travisano:Science:1995,weinreich:Evolution:2005a,travisano:Evolution:2013}.

Epistasis is any deviation from the additive effects of alleles at different loci \cite{fisher:TRE:1918}.
Epistasis gives rise to rugged fitness landscapes which have been found to be quite common in experimental observations in a variety of model systems \cite{jain:Genetics:2007,szendro:PNAS:2013}. 
In particular, reciprocal \textit{sign epistasis} is a necessary condition for having a rugged fitness landscape \cite{poelwijk:Nature:2007}. 
While in \textit{magnitude epistasis} the fitness always increases (or decreases) with every additional mutation in a non-additive manner, 
in \textit{sign epistasis}, however, valleys appear in the fitness landscape. 
A certain mutation might have a lower fitness than the previous state although it leads to higher fitness eventually. 
In such a case not all paths in the fitness landscape might be accessible by the population \cite{weinreich:Science:2006}. 
Comparing experimental systems to theoretical predictions made on the basis of the underlying fitness landscape helps elucidate the role of microscopic properties of the system in determining the macroscopic evolutionary trajectory. 
The details of the process such as the mutation rate, fitnesses of individual states and the global population size act as constraints on the accessibility of paths \cite{szendro:PNAS:2013}.
Using the assumption of strong selection and weak mutation rates (SSWM), the system advances on the fitness landscape in a stepwise fashion.
This automatically limits the possible number of adaptive paths \cite{weinreich:Evolution:2005a}.

Evolutionary predictability and the speed of the dynamics is not only determined by the molecular constraints of fitness and mutation rate but also by population dynamics \cite{poelwijk:Nature:2007}.
Theoretical explorations often assume a fixed population size starting at one node of the fitness landscape and its movement is tracked over the course of time. 
Increasing the population size, or the mutation rate, we observe the phenomenon of clonal interference \cite{weinreich:Science:2006, park:PNAS:2007}. 
This occurs when a second step mutant arises in a population even when the first step mutation is not fixed. 
In other words, the SSWM assumption is no longer valid. 
Clonal interference has been extensively explored experimentally \cite{imhof:PNAS:2001,elena:NRG:2003,hegreness:Science:2006} as well as theoretically \cite{gerrish:Genetica:1998,iwasa:Genetics:2004,weinreich:Evolution:2005b,desai:CB:2007,park:PNAS:2007,gokhale:JTB:2009,weissman:TPB:2009}.
This phenomenon removes the limit on the accessibility of non-adaptive trajectories. 
If the fitnesses and mutation rates align to particular conditions, i.e. the mutation rates also underlie epistatic interactions, then such valley crossings might be faster than adaptive trajectories \cite{gokhale:JTB:2009,lynch:MBE:2010}. 

Populations in real systems are finite and their size can undergo fluctuations which can lead to possible extinction events.
Together with the phenomena of clonal interference and epistatic interactions between mutations (correlated rugged fitness landscapes), predicting evolution through a given fitness landscape seems like an impossible task.
Herein we develop a general methodology for predicting the most probable path in a fitness landscape with epistatic interactions in a multi-dimensional fitness landscape.
To reflect a realistic scenario we use a multi-type branching process (e.g. \cite{haccou:book:2005}) to drop the assumption of a constant population size. 
For presentation purposes we limit ourselves to systems without back mutations.
The model in its full generality is free of this assumption, although it is unclear how to define pathways when back mutations are allowed (see Supplementary Information for a detailed explanation).
To introduce the framework we begin with a simple model in which the wild type can have two independent mutations leading to the fittest type. 
Then we increase the number of mutational events it takes to get to the corresponding type leading to a generalization of the methodology.
We briefly mention an application of this approach by linking it to a cancer initiation model \cite{bauer:JTB:2014} showing how mutational epistasis changes the path probabilities.
Finally we provide an outline on how to extend the model to a general system where different mutations need to be acquired to reach the final mutant.

\section*{Methods and Results}

\subsection*{Probability Generating Function}

For our methodology, we are making use of extinction probabilities, more specifically the probability for different types to be present or not to be present.
In a branching process this probability can be recursively obtained using \textit{probability generating functions} (PGF).
Since the relation between PGFs and the probability for a type to be present is the main tool we are using, we devote this subsection to giving a short overview about this correlation, although it is rather technical and well known (e.g.\ \cite{haccou:book:2005, kimmel:book:2002}).

The probability generating function (PGF) in discrete time for a one-type process is in general defined as
\begin{align}
 f(s) = \sum_{k=0}^\infty p_k s^k,
\end{align}
where $k$ denotes the number of offspring and $p_k$ represents the probability of having $k$ offspring (the focal individual dies in this context) \cite{haccou:book:2005}.
For many biological processes, for example cell multiplication, it makes sense to only consider offspring numbers of 0 (death), 1 (nothing happens), and 2 (cell division).
But in other biological systems it makes sense to consider many offspring at once, for example reproduction via numerous seeds in plants.
Our analysis is not restricted to any particular offspring distribution.
However, for the sake of simplicity, we restrict our example to the so called \textit{binary splitting}, i.e.\ either two or no offspring.
The use of the argument $s$ is not obvious at this point.
If we set $s$ equal to $0$, the probability generating function reduces to $f(0)=p_0$, which is the extinction probability for a population of one individual in one time step.
Since all individuals behave independently, $f(0)^N = p_0^N$ is the extinction probability for a population of size $N$ in one time step.
Now looking at the extinction probability within two time steps, we note that with probability $p_2$ we would have two individuals in the next time step originating from one individual.
Hence, the extinction probability for a single individual within two time steps is,
\begin{align}
 p_0 + p_2 p_0^2 = f(f(0)) = f^{ \circ (2)}(0),
\end{align}
and that of population with $N$ individuals is,
\begin{align}
 \left( p_0 + p_2 p_0^2\right) ^N = \left( f(f(0))\right) ^N = \left( f^{ \circ (2)}(0) \right)^N.
\end{align}
Continuing for further time steps, we see that $ f^{\circ (t)}(0)$ is the extinction probability for the system within $t$ time steps.

As of now we assumed that individuals reproduce clonally i.e. giving rise to the same type.
Now we continue investigating the extinction probability for a two-type process. 
Let us think of the two types $A$ and $B$, where an $A$ individual can produce any number of $A$ or $B$ individuals, and respectively for $B$. Then the general PGFs if the process starts with one type $A$ or one type $B$ individual are defined as
\begin{align}
 f_A(s_A, s_B) = \sum_{k_A=0}^\infty \sum_{k_B=0}^\infty p^A_{k_A,k_B} s_A^{k_A}s_B^{k_B}, \\
 f_B(s_A, s_B) = \sum_{k_A=0}^\infty \sum_{k_B=0}^\infty p^B_{k_A,k_B} s_A^{k_A}s_B^{k_B},
\end{align}
where $p^A_{k_A,k_B}$ $(p^B_{k_A,k_B})$ denotes the probability of one $A$ ($B$) individual producing $k_A$ $A$ and $k_B$ $B$ individuals in the next time step.
Let us try to recover the extinction probability as for the one-type process.
If we set both $s_A$ and $s_B$ equal to zero and assume that we start with one $A$ individual, we obtain a similar result as above for the total extinction probability
\begin{align}
 f_A(0, 0) =p^A_{0,0}.
\end{align}
Oftentimes, one is rather interested in the extinction, or non-presence, of just one particular type.
Let us for example assume we are only interested in the presence of $B$ individuals.
The probability of having no $B$ individuals in time step 1 is the sum over all probabilities, where no $B$ offspring is being produced $\sum_{k_A=0}^\infty p^{A(B)}_{k_A,0}  = f_{A(B)}(1, 0)$, starting with one $A$ ($B$) individual.
Now looking at the probability of having no $B$ individuals in time step 2, we need to account for the probability of having $k_A$ A and $k_B$ B individuals being produced in the first time step. This leads to
\begin{align}
 \sum_{k_A=0}^\infty \sum_{k_B=0}^\infty p^A_{k_A,k_B} f_A(1,0)^{k_A} f_B(1, 0)^{k_B} = f_A(f_A(1,0), f_B(1, 0)) =: f_A^{\circ (2)}(1,0).
\end{align}
Continuing this procedure and analogous to the one-type process, the probability of having no $B$ individual in time $t$ is $f_A^{\circ (t)}(1,0)$.

In a similar fashion this procedure can be extended to a multi-type process with an arbitrary number of types.
For further information and detailed insights into extinction of branching processes we refer to \cite{haccou:book:2005} and \cite{kimmel:book:2002}.

\subsection*{Two dimensional fitness landscape}

\begin{figure}[htp]
\begin{center}
 \includegraphics[width=0.3\textwidth]{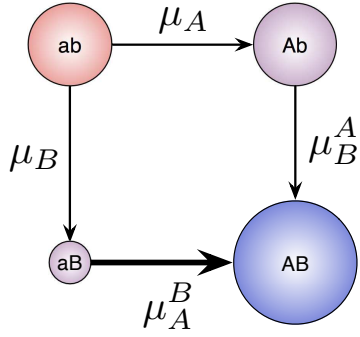}
\caption{
\label{fig:minmodelfitnesslandscape}
\textbf{Mutational pathways for a system with two loci.} There are two different pathways to reach the final mutant.  
Fitness is represented by the size of the circles denoting the types.
Thus the wildtype $ab$ and $Ab$ have a similar fitness whereas $AB$ has a significantly greater fitness compared to the wildtype while $aB$ is much less fit than the wildtype.
When all mutation rates are the same, the pathway via $aB$ would be not adaptive, since this type has a low fitness. If the mutation rate $\mu_A^B$ is large enough, especially if $\mu_A^B \gg \mu_A$ (indicated by the thick arrow), this pathway becomes accessible.}
\end{center}
\end{figure}

We begin with a minimal fitness landscape. 
Envision a wildtype $ab$ which can mutate at the two loci to $A$ and $B$, respectively. 
With both mutations, the system is in the final state of $AB$.
In such a system there are two different paths as illustrated in Figure \ref{fig:minmodelfitnesslandscape}.
Traditionally, epistatic models are discussed in terms of different fitness values, whereas the mutation rates stay the same \cite{poelwijk:Nature:2007,szendro:PNAS:2013}. 
Exemplarily the fitness landscape for a system with \textit{sign epistasis} is shown in Figure \ref{fig:minmodelfitnesslandscape}.
In such a system where the mutation rates stay the same, i.e. $\mu_A=\mu_A^{B}$ and $\mu_{B}=\mu_{B}^{A}$, it is clear that the path via $Ab$ is the most probable one. 
However, if the mutation rates change, e.g. $\mu_{A}^{B}\gg \mu_{A}$, also the path via $aB$ can become accessible.
Changing mutation rates amounts to including epistasis in the mutational landscape in addition to epistasis in the fitness landscape \cite{sasaki:JTB:2003}.

For the four types of the above model, we need to consider four different PGFs, one for each type
 \begin{eqnarray}
  f_{ab}  (s_{ab},s_{Ab}, s_{aB}, s_{AB}) &=&  d_{ab} + b_{ab}((1-\mu_A-\mu_B)s_{ab} + \mu_A s_{Ab} + \mu_B s_{aB})^2,
 \nonumber \\
  f_{Ab}  (s_{ab},s_{Ab}, s_{aB}, s_{AB}) &=&  d_{Ab} + b_{Ab}((1-\mu_B^A)s_{Ab} + \mu_B^A s_{AB})^2,  \nonumber \\
  f_{aB} (s_{ab},s_{Ab}, s_{aB}, s_{AB}) &=& d_{aB} + b_{aB}((1-\mu_A^B)s_{aB} + \mu_A^B s_{AB})^2, \nonumber \\
  f_{AB} (s_{ab},s_{Ab}, s_{aB}, s_{AB})  &=& d_{AB} + b_{AB} s_{AB}^2,
 \end{eqnarray}
where $b_i$ and $d_i$ are the birth and death probabilities of type $i$. 
The exponent of $2$ arises from a branching process with binary splitting.
The arguments $s_{ab}, \ldots, s_{AB}$ correspond to extinction probabilities of the respective type as discussed above.
The functions $f_{i}$ correspond to the extinction probability of the whole process given that the process starts with a single individual of type $i$. 
The PGF $f_{i}$ at time $t$ is recursively calculated as
\begin{eqnarray}
 f_{i}^{(t)}(s_{ab},s_{Ab}, s_{aB}, s_{AB})  = f_{i}(f_{ab}^{(t-1)},f_{Ab}^{(t-1)}, f_{aB}^{(t-1)}, f_{AB}^{(t-1)}). 
\end{eqnarray}

\subsection*{Time Distribution}
Using the generating functions we now approach the extinction time distribution of the binary branching process.
Particularly starting with 1 wild type individual, the probability of having no ${AB}$-individual at time $t$ is $f_{ab}^{(t)}(1,1,1,0)=:f(t)$.
Thus the probability of having at least 1 ${AB}$-individual at time $t$ is $1-f(t)$.
The probability, that at least 1 ${AB}$-individual appears \textit{exactly} at time $t$ is the probability, that there is an ${AB}$-individual at $t$ minus the probability that there was already one at time $t-1$:
\begin{equation} \label{eq:timedistri}
\tau(t)=(1-f(t)) - (1 - f(t-1)) = f(t-1)-f(t).
\end{equation}
Starting with $N$ wild type individuals the probability that there are no $AB$-individual at time $t$ is then $f(t)^N$. 
This leads to the time distribution as,
\begin{equation}
 \tau(t) = f^N(t-1) - f^N(t).
\end{equation}

\begin{figure}[htp]
\begin{center}
 \includegraphics[width=0.6\columnwidth]{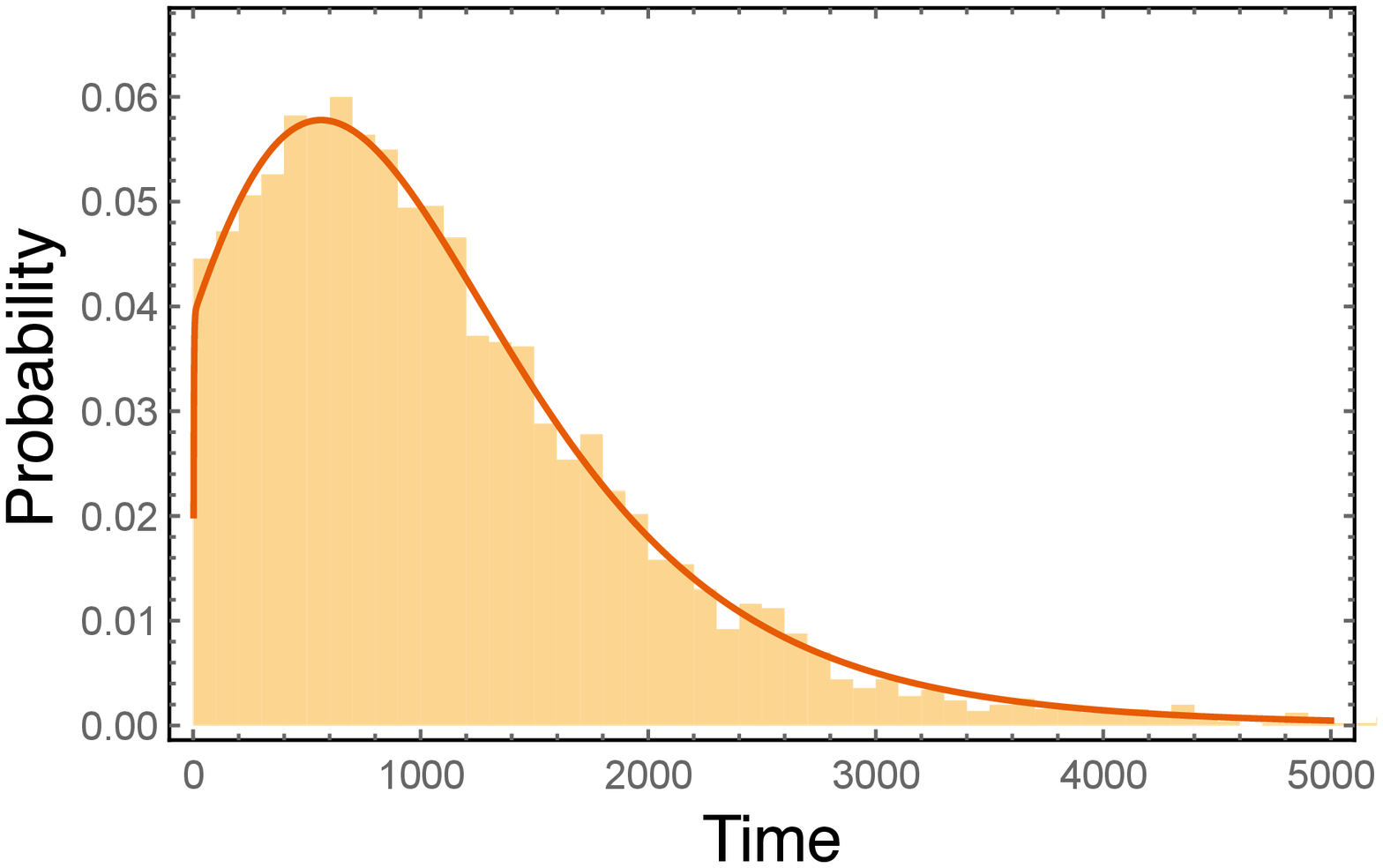}
\caption{\textbf{Time distribution of reaching the final mutant for a four type fitness landscape as in Fig.~\ref{fig:minmodelfitnesslandscape}.}
Solid line represents the recursive solution and the bars represent $5000$ simulations. 
The parameters are: \newline
Death probabilities: $d_{ab}=0.5, d_{Ab}=0.49995, d_{aB}=2/3, d_{AB}=0.25$. Birth probabilities are 1 minus the corresponding death probability. Mutation probabilities are $\mu_B=\mu_{B}^A=2\cdot 10^{-6}$, $\mu_{A}=2\cdot 10^{-5}$, $\mu_{A}^B=0.005$. Population size in the beginning: $N=30000$.
\label{fig:2x2times}
}
\end{center}
\end{figure}

However, the arising ${AB}$ should start a lineage that does not die out. 
Hence we are interested in the probability of having a \textit{successful} ${AB}$-individual.
To calculate this we use the known extinction probability of an $AB$-individual in place of $s_{AB}$.
The probability of an $AB$-individual going extinct is its death probability divided by its birth probability $e_{AB}:=d_{AB}/b_{AB}$ \cite{athreya:book:1972}. 
The modified PGFs for this purpose then read as
\begin{eqnarray}
\label{eq:pgfext}
 f_{ab}(s_{ab},s_{Ab}, s_{aB}) &=&  d_{ab} + b_{ab}((1-\mu_A-\mu_B)s_{ab} + \mu_A s_{Ab} + \mu_B s_{aB})^2,
 \nonumber \\
  f_{Ab}(s_{ab},s_{Ab}, s_{aB}) &=& d_{Ab} + b_{Ab}((1-\mu_B^A)s_{Ab} + \mu_B^A e_{AB})^2,  \nonumber \\
  f_{aB}(s_{ab},s_{Ab}, s_{aB}) &=& d_{aB} + b_{aB}((1-\mu_A^B)s_{aB} + \mu_A^B e_{AB})^2.
\end{eqnarray}
Note, that the PGF for the final mutant type is not necessary anymore.
We can now calculate the time distribution until the first \textit{successful} mutant appears the same way as described above.
Figure \ref{fig:2x2times} shows the perfect agreement between the recursive solution and $5000$ simulations.
The parameters, specified in the Figure \ref{fig:2x2times}'s caption, are entirely arbitrarily chosen to reflect an epistatic fitness landscape as sketched in Figure \ref{fig:minmodelfitnesslandscape}.
The reason we chose a very slightly advantageous fitness for the type $Ab$-individuals is solely to stress the fact, that this method holds for any fitness values, not only if some are restricted, for example to being neutral.

For a three-type continuous time branching process, as in $ A \stackrel{\mu_{B}}{\longrightarrow} B \stackrel{\mu_{C}}{\longrightarrow} C$, the time distribution was computed in \cite{bozic:ELIFE:2013}.
This was done using the analytical solution of the probability generating function for the two-type process $A \stackrel{\mu_{B}}{\longrightarrow} B$ \cite{antal:JSM:2011} and the fact, that in continuous time mutations follow a Poisson distribution. 
Adding a second intermediate type, e.g. $B_2$, would also give such a process but immediately results in unwieldy analytical calculations.

\subsection*{Path Probabilities}

In the current example there are two possible paths by which the wildtype can reach the final mutant $AB$,
either $ab$ $\rightarrow$ $Ab$ $\rightarrow$ $AB$ or $ab$ $\rightarrow$ $aB$ $\rightarrow$ $AB$. 
Experimental evidence shows that not all paths are equally probable \cite{lee:JMoE:1997,weinreich:Science:2006}.
Beginning with $ab$ then what is the probability of the first $AB$ mutant arising via either path and how long does it take for the different pathways?
 
The probability, that the first mutant arises exactly at time $t$ via pathway Ab is (derived in the SI),
\begin{equation}
 \rho_{Ab}(t) = f^N(t-1) - (\bar{f}^{(Ab)}(t))^N,
 \label{eq:pathprob}
\end{equation}
where $\bar{f}^{(Ab)}(t)$ is defined in the Supporting Information (SI) and is being computed in a similar fashion as $f(t)$.
The total probability for this path $\varrho_{Ab}$ is then the summation of $\rho_{Ab}(t)$
\begin{equation}
\varrho_{Ab} = \sum_{t=1}^\infty \rho_{Ab}(t).
\end{equation}
Computationally the sum would go up to a $t_{max}$, where $f^{(Ab)}(t_{max}-1)-f^{(Ab)}(t_{max})<\epsilon$ (where usually \textit{machine epsilon} is chosen as $\epsilon$).
The total extinction probability of a multi-type branching process is determined by the smallest fixed point $\mathbf{s}^* = (s_{ab}^*, s_{Ab}^*, s_{aB}^*, s_{AB}^* )$ of the probability generating functions $\mathbf{f}(\mathbf{s}^*) = \mathbf{s}^*$, where $s_{ab}^*$ is the extinction probability, if the process starts with one $ab$-individual \cite{haccou:book:2005}. 
Nevertheless those total extinction probabilities are not suitable for the question, via which path the first successful $AB$-mutant arises. 
The problem lies in the time; the pathway via $Ab$ for example could have a very low extinction probability whereas the pathway via $aB$ might have an extinction probability of $1/2$. 
Intuitively one would expect the path via $Ab$ to be more frequent. 
However, if the path via $aB$ is much faster (e.g. due to $\mu_A^B \gg \mu_B^A$) one would actually find that each path happens with probability that approaches 1/2. 
Therefore, it is important to do the recursive analysis to include the probability, that a successful mutant did not arise through any other path beforehand.

\begin{figure}[htp]
\begin{center}
 \includegraphics[width=0.6\columnwidth]{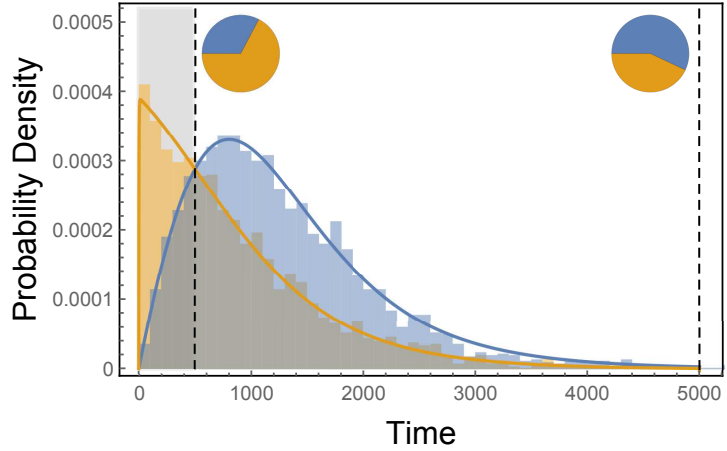}
\caption{
\label{fig:2x2paths}
\textbf{Probability distribution for the different pathways.}
Orange represents the pathway via $aB$ and blue the pathway via $Ab$.
The bars are the results of simulations, the solid lines depict the computed results.
In the pie charts the distribution of the pathways are illustrated up to 500 time steps (shaded area, left pie chart) and up to 5000 time steps (right pie chart). 
Stopping after a few lineages have reached the final mutant might lead to a false distribution:
The other pathway might just need longer, but have a smaller extinction probability.
The parameters are: 
Death probabilities: $d_{ab}=0.5, d_{aB}=2/3, d_{Ab}=0.49995, d_{AB}=0.25$.
Birth probabilities are 1 minus the corresponding death probability.
Mutation probabilities are $\mu_B=\mu_{B}^A=2\cdot 10^{-6}$, $\mu_{A}=2\cdot 10^{-5}$, $\mu_{A}^B=0.005$. Initial Population size is $N=30000$.}
\end{center}
\end{figure}

Figure \ref{fig:2x2paths} shows the probability densities for the different pathways of the minimal model. 
Interestingly, the pathway via aB is predominantly prominent in the beginning but overall less likely. 
Hence if experiments are stopped after a short time interval then they might provide conclusions which can be upended by looking at the experiments at a later time point.

\subsection*{Multiple mutations in two dimensions}
\begin{figure}[htp]
\begin{center}
 \includegraphics[width=0.5\columnwidth]{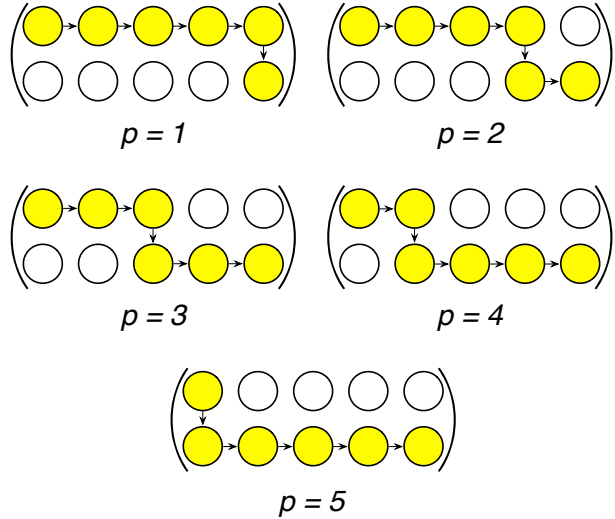}
 \caption{
 \label{fig:path_numb} 
\textbf{Exemplary numbering of the different mutational pathways in a system with $m=4$ mutations for type $A$ and $n=1$ mutation for $B$.}}
 \end{center}
\end{figure}

\begin{figure}[htp]
\centering
 \includegraphics[width=1.\columnwidth]{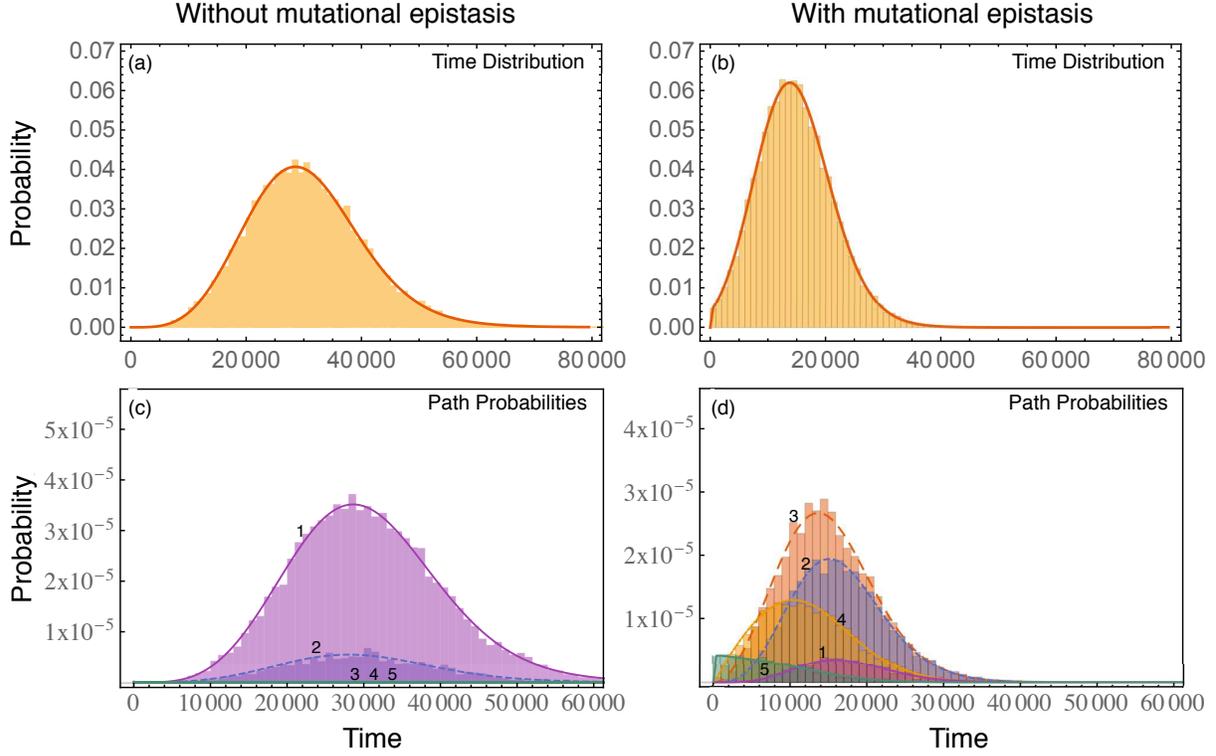}
\caption{ \label{fig:burkitt_plots}
\textbf{Comparison between the path probability distributions of a minimal Burkitt Lymphoma model. }
Top: Time distributions for the model (a) without epistatic effects on mutation probabilities and (b) with mutational epistasis. The probability to obtain an $A$ mutation is 100 times higher, if the $B$ mutation is present in that individual. 
Bottom: In (c) the path probabilities for the model without epistatic effects on mutations are illustrated, whereas in (d) the mutation probability is again increased by 100 for an $A$ mutation if the $B$ mutation is present. 
Pathway 1 corresponds the the mutational pathway, where first all necessary extra mutations have to be acquired, and the $B$ mutates last. Pathway 2 denotes the pathway, where 3 of 4 extra mutations have been obtained, then the $B$ mutation happens, and at last the final extra mutation is acquired. Respectively for the other pathways (cf. Figure \ref{fig:path_numb}). 
The parameters are the same as in \cite{bauer:JTB:2014}: 
The birth probability for an individual with $j$ passenger mutations and without the $B$ mutation is $b_{0,j}=0.5 (1+10^{-5})^j$, and with the $B$ mutation $b_{1,j}=\frac{1.05}{2.2}\cdot 1.015^j$.
The mutation probability for the $B$ mutation is $\mu_D=5\cdot 10^{-6}$, for an $A$ mutation without the $B$ mutation being present $\mu_{P}=2\cdot 10^{-5}$, and with the $B$ mutation being present (only necessary for (b) and (d)) $\mu_{D}^P=2\cdot 10^{-5}$.
The population size in the beginning is $N=500000$.
}
\end{figure}

In the earlier model the wildtype had two possible mutations $a \rightarrow A$ and $b \rightarrow B$. 
It is possible, that $a$ to $A$ and $b$ to $B$ are a multi-step process.
Hence we can assume that it takes $m$ mutations to go from $a$ to $A$ and $n$ to go from $b$ to $B$.
Hence for $m=n=1$ we recover the simple model as discussed above. 
The calculation of the time distribution can be directly transferred from the simple model by including all necessary probability generating functions for all available types.
Increasing the length of the dimensions has a direct impact on the number of paths leading from the wildtype to the final mutant.
In particular there are $N = \binom{m+n}{m}$ possible paths.
Assuming in general $m$ mutations in the $A$ dimension and $n$ in the $B$ dimension we enumerate the paths as follows. 
Path 1 is the path where at first all $A$ mutations and subsequently all $B$ mutations happen.
Path 2 is the path where all but one $A$ mutations happen first, then one $B$, then the last $A$, and finally all other $B$ mutations.
Figure \ref{fig:path_numb} shows the different paths for a system with four mutations for type $A$ and one mutation for type $B$.
Thus calculating the path probability for any particular path $p$ now takes the form,
\begin{equation}
\rho_p(t) = f^N(t-1) - \left(\bar{f}^{(p)}(t) \right)^N,
\end{equation}
where $f(t)$ is the probability generating function as in Eq.\ A.2 and $\bar{f}^{(p)}$ is defined analogously to Eq.\ A.9 in the SI
\begin{eqnarray}\label{eq:path2D}
\bar{f}^{(p)}(t)&:= & \bar{f}_{p_0}^{\circ (t)} \left( \underbrace{1, 1, \ldots, 1}_{\substack{m+n}}, \frac{d_{m,n}}{b_{m,n}}, 1, \ldots, 1 \right) \nonumber \\
                &=& \bar{f}_{p_0} \left( \bar{f}_{p_0}^{\circ (t-1)}, \bar{f}_{p_1}^{\circ (t-1)}, \ldots, \frac{d_{m,n}}{b_{m,n}}, \bar{f}_{q_1}^{\circ (t-2)}, \ldots, \bar{f}_{q_{mn}}^{\circ (t-2)} \right).
\end{eqnarray}
Here, the probability generating functions with a $p$ index belong to types along the regarded path (which in total are $m+n+1$ without back mutations, beginning at $0$, with which we always label the subindex for the wild type).
Accordingly, probability generating functions with a $q$ index are associated with types, that do not belong to the respective path (which are in total $m \times n$).
The probability generating function for the final mutant type is again replaced by the extinction probability of this type.
We use our framework with this extension on the cancer initiation model proposed in \cite{bauer:JTB:2014}. Therein a model with several mutational steps to reach state $A$ and one mutational step for state $B$ is analyzed (cf.\ Fig. \ref{fig:path_numb}). The direct change in fitness for the $A$ mutations is (nearly) zero, and the $B$ mutation alone is even deleterious. However, if an individual obtains all $A$ mutations and the $B$ mutation, the fitness is enhanced which in the model leads to rapid proliferation.
Here, we provide an example on how the path probabilities change, when epistasis is not just in the fitness landscape but in the mutational landscape as well.
Figure \ref{fig:burkitt_plots} compares the path probability distributions with and without epistasis in the mutational landscape. 
The fitness values, the birth and death probabilities respectively, as well as the ``nonepistatic'' mutation probabilities, are the same as in \cite{bauer:JTB:2014}.

\subsection*{Multi dimensional fitness landscapes}
The cancer landscape discussed above is a two dimensional system.
In principle it is possible to extend this approach to higher dimensions. 
For fitness landscapes of higher orders \cite{weinreich:Science:2006,khan:Science:2011} it is still possible to write down the system of probability generating functions and apply the approach explained here.
The concept remains the same. For each type the probability generating functions are needed except for the final mutant type, here only the extinction probability is necessary (SI). 
Finally the probability generating function for the wild type needs to be recursively calculated for the time distribution. 
For the path probabilities the probability generating functions related to types not along the considered path again are one time step behind, similar as in Eq.\ \ref{eq:path2D}.
However for these experimental fitness landscapes while we can get accurate data elucidating the fitness landscape, the mutational landscape is usually hard to determine.

\section*{Discussion}

We have presented a theoretical framework to study mutational pathways in epistatic systems.
The crucial part is that in our analysis epistasis affects not only fitness (i.e. proliferation and death rates) but also mutation rates.
Hereby we could show, that pathways become accessible, that without mutational epistatic effects are mostly unlikely to happen (cf.\ e.g.\ Figure \ref{fig:burkitt_plots}). Our analysis is based on multi-type branching processes and hence it does not rely on the assumption of a constant population size. 

While we have focused on a fairly simple system with a fitness landscape with a single peak, the approach can be extended to a rugged fitness landscape.
Moreover, if back mutations are involved, one can still calculate the time distribution, although pathways are not clearly defined in a system with back mutations anymore (see SI).
Furthermore in the current scenario in each time step the individuals could replicate or die. 
In addition we could have a resting probability where the individuals remain in the same state with a certain probability.
Such complicated scenarios can be incorporated in our framework as well (SI).
The computations can be precisely represented in analytic terms and need to be solved recursively.

We apply our framework to a cancer model  including mutational epistasis \cite{bauer:JTB:2014} and show how the path probabilities are altered by it.
Mutational epistasis can thus lead to heterogeneity in the density of different mutant types between different age groups as reaching the final mutant early is only possible by one mutational pathway which is not possible at later time points.

As shown here the mutational landscape can undermine the current predictions based solely on fitness landscapes.
Just like in long term evolution, experimental as well as theoretical approaches ought to be balanced between studying effects of selection \textit{and} the strengths of mutations.
The theoretical analysis based on the approach explained here helps in understanding the importance of mutational epistasis, even though the computations have to be solved recursively.
In particular, it makes analyzing the fitness and mutational landscapes more interactive, since long-lasting simulations are not necessary any more.
 \\

\noindent\textbf{Acknowledgements}\\
We thank Laura Hindersin and Arne Traulsen for providing constructive comments on the manuscript.
Funding from the Max Planck Society, the New Zealand Institute for Advanced Study and the DFG Priority Programme 1590 Probabilistic Structures in Evolution (Grant GO2270/1-1) is gratefully acknowledged.

\section*{Additional Information}

\textbf{Contributions}\\
B.B. did the mathematical analysis and performed simulations. B.B. and C.S.G. developed the recursive algorithm and wrote the manuscript.\\
\textbf{Competing financial interests}\\
The authors declare no competing financial interests.

\newpage

\section*{Supporting Information: Repeatability of evolution on epistatic landscapes}

\subsection*{General Probability Generating Functions}

\begin{figure}[htp]
\begin{center}
 \includegraphics[scale=0.5]{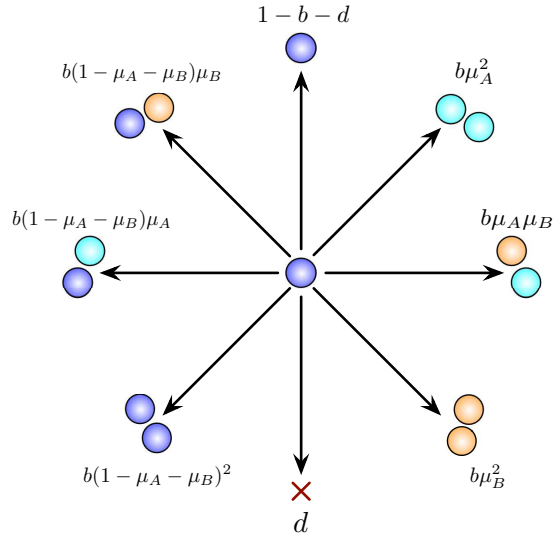}
 \end{center}
\caption{
\label{fig:genpfg}
\textbf{Process described by the general pgf.} An individual can either die, proliferate, or neither and just live. If it proliferates the offspring can mutate.
In case of including back mutations additional mutation terms appear leading as in Eq.~\eqref{eq:pgfwithbm}. 
}
\end{figure}

In the main text we considered only the case where each individual has to die or divide in every time step. 
Here we relax this assumption and consider a more realistic scenario where only some individuals proliferate or die, whereas others do not take any action at all (Fig.~\ref{fig:genpfg}). 
Then, the probability generating functions for the four types: wild type, individuals with mutation $A$, individuals with mutation $B$, and individuals with both mutations are defined as
 \begin{align}
  f_{ab}(s_{ab},s_{Ab}, s_{aB}, s_{AB}) & = d_{ab} + (1-b_{ab}-d_{ab})s_{ab} + b_{ab}((1-\mu_A-\mu_B)s_{ab} + \mu_A s_{Ab} + \mu_B s_{aB})^2, \nonumber \\
  f_{Ab}(s_{ab},s_{Ab}, s_{aB}, s_{AB}) & = d_{Ab} + (1-b_{Ab}-d_{Ab})s_{Ab} + b_{Ab}((1-\mu_B^A)s_{Ab} + \mu_B^A s_{AB})^2, \nonumber \\
  f_{aB}(s_{ab},s_{Ab}, s_{aB}, s_{AB}) & = d_{aB} + (1-b_{aB}-d_{aB})s_{aB} + b_{aB}((1-\mu_A)s_{aB} + \mu_A s_{AB})^2, \nonumber \\
  f_{AB}(s_{ab},s_{Ab}, s_{aB}, s_{AB}) & = d_{AB} + (1-b_{AB}-d_{AB})s_{AB} + b_{AB} s_{AB}^2.
  \label{eq:pgfext}
 \end{align}
The functions are similar to the scenario of binary splitting (cf. Eq. 1 in the main text). There is only one term added: $(1-b_i-d_i)s_i, i \in \{ab, Ab, aB, AB\}$ which denotes the case of the individual neither dividing nor dying.
To make the model even more realistic one could also include back mutations,
\begin{align}
  f_{ab}(s_{ab},s_{Ab}, s_{aB}, s_{AB}) & = d_{ab} + (1-b_{ab}-d_{ab})s_{ab} + b_{ab}((1-\mu_A-\mu_B)s_{ab} + \mu_A s_{Ab} + \mu_B s_{aB})^2 \nonumber \\
  f_{Ab}(s_{ab},s_{Ab}, s_{aB}, s_{AB}) & = d_{Ab} + (1-b_{Ab}-d_{Ab})s_{Ab} + b_{Ab}((1-\mu_{ab}^A-\mu_B^A)s_{Ab} + \mu_{ab}^A s_{ab} + \mu_B^A s_{AB})^2 \nonumber \\
  f_{aB}(s_{ab},s_{Ab}, s_{aB}, s_{AB}) & = d_{aB} + (1-b_{aB}-d_{aB})s_{aB} + b_{aB}((1-\mu_{ab}^B-\mu_A^B)s_{aB} + \mu_{ab}^B s_{ab} + \mu_A^B s_{AB})^2 \nonumber \\
  f_{AB}(s_{ab},s_{Ab}, s_{aB}, s_{AB}) & = d_{AB} + (1-b_{AB}-d_{AB})s_{AB} \nonumber\\
                               &\ \ \ \ + b_{AB} \left((1-\mu_A^{AB}-\mu_B^{AB})s_{AB} + \mu_A^{AB} s_{Ab} + \mu_B^{AB} s_{aB}\right)^2 \nonumber \\
\label{eq:pgfwithbm}
\end{align}
If the fitness landscape is rugged, i.e. having multiple local optima, they would be inaccessible from certain ``downstream" directions if back mutations are not allowed. 
Hence allowing back mutations, allows to have a rugged fitness landscape with local optima accessible from multiple directions.
The probability generating functions seem more complex, but the principle of the computation as discussed in the main text does not change at all. 

\subsection*{Time distribution}

Here, we give a more detailed description on how to calculate the time distribution for the minimal model with four types, and two paths, but with back mutations.
\begin{enumerate}
  \item Calculate the extinction probability of the final mutant type $AB$ as in \cite{athreya:book:1972}
  \begin{align}
    e_{AB}=\frac{d_{AB}+b_{AB} \left( \mu_A^{AB} + \mu_B^{AB}\right)^2}{b_{AB}(1-\mu_A^{AB}-\mu_B^{AB})^2}.
  \end{align}
Note, that without back mutations the extinction probability reduces to $e_{AB}=\frac{d_{AB}}{b_{AB}}$ as in the main text.
  
  \item Until some $t_{max}$ calculate recursively
  \begin{align} \label{eq:timedistri}
    f_{AB}^{\circ (t)} & = d_{AB} + (1-b_{AB}-d_{AB})f_{AB}^{\circ (t-1)} \nonumber \\
                               &+ b_{AB} \left((1-\mu_A^{AB}-\mu_B^{AB})f_{AB}^{\circ (t-1)} + \mu_A^{AB} f_{Ab}^{\circ (t-1)} + \mu_B^{AB} f_{aB}^{\circ (t-1)}\right)^2 , \nonumber \\
    f_{aB}^{\circ (t)} & = d_{aB} + (1-b_{aB}-d_{aB})f_{aB}^{\circ (t-1)} + b_{aB}\left((1-\mu_A^B - \mu_{aB}^B) f_{aB}^{\circ (t-1)} + \mu_{aB}^B f_{ab}^{\circ (t-1)} + \mu_A^B f_{AB}^{\circ (t-1)}\right)^2, \nonumber \\
    f_{Ab}^{\circ (t)} & = d_{Ab} + (1-b_{Ab}-d_{Ab})f_{Ab}^{\circ (t-1)} + b_{Ab}\left((1-\mu_B^A - \mu_{aB}^A) f_{Ab}^{\circ (t-1)} + \mu_{aB}^A f_{ab}^{\circ (t-1)} + \mu_B^A f_{AB}^{\circ (t-1)}\right)^2, \\
    f(t)&:=f_{ab}^{\circ (t)} = d_{ab} + (1-b_{ab}-d_{ab})f_{ab}^{\circ (t-1)} \nonumber \\
                    & + b_{ab}\left((1-\mu_A-\mu_B)f_{ab}^{\circ (t-1)} + \mu_A f_{Ab}^{\circ (t-1)} + \mu_B f_{aB}^{\circ (t-1)} \nonumber\right)^2
  \end{align}
  where $f_{aB}^{\circ (0)} = f_{Ab}^{\circ (0)} = f_{ab}^{\circ (0)} = 1$ and $f_{AB}^{\circ (0)} = e_{AB}$.
  Note, that without back mutations these functions would not be coupled anymore and one can first calculate $f_{Ab}^t$ and $f_{aB}^t$ for all $t$, since those functions would not depend on $f_{ab}$.
  Moreover, $f_{AB}^{\circ (t)}$ would be equal to $e_{AB}$ $\forall$ $t$. Hence, one would not need to recursively calculate $f_{AB}^{\circ (t)}$. 
  However, the complexity does not change.
  
  \item The probability to get the final, successful $AB$ mutant, i.e. an individual that produces a lineage that does not die out again, exactly at time $t$ is
  \begin{align}
    \tau(t)=f^N(t-1)-f^N(t).
  \end{align}
  where $N$ is the number of individuals in the beginning.
  Calculating this for all $t \in \{ 0, \ldots, t_{max}\}$ we obtain the time distribution.
\end{enumerate}

\subsection*{Single-Path time distribution}
Here, we explain the computation of the probability distribution of the pathway via type $Ab$ exemplarily.
Allowing back mutations it is unclear how to specify different mutational pathways.
For instance for the pathway $ab \rightarrow aB \rightarrow ab \rightarrow Ab \rightarrow AB$ it is obscure to say via which type the final mutant has been reached.
Obviously the final mutant has been reached via type $Ab$, but it might be necessary for the population to first reach type $aB$. 
Hence, $aB$ might play a vital role for reaching $AB$, too. For this reason we neglect back mutations in the computation of the path probabilities, thus guaranteeing clear distinguishable pathways.

Let Ab($t$) (aB($t$)) denote the random variable, that there is an AB mutant until time $t$ via pathway Ab (aB).
Thus, $\neg$Ab($t$) corresponds to the random variable, that there is no AB mutant until time $t$ vial pathway Ab.
Then the probability, that the first mutant arises exactly at time $t$ via pathway Ab (i.e. not via pathway aB beforehand) is
 \begin{align} \label{eq:probs}
 \rho_{Ab}(t)=& P(Ab(t) \cap \neg Ab(t-1) \cap \neg aB(t-1)) \nonumber \\ 
 = & P(\neg Ab(t-1)  \cap \neg aB(t-1))-P(\neg Ab(t)  \cap \neg aB(t-1)).
 \end{align}
The first term is calculated by the pgf as in Eq.\ \eqref{eq:pgfext}.
For the second term however, the time points for the different pathways are different.
Let us derive a recursive function for this second term at this point.
To do so, let us first consider the extinction probability for the subprocess of $Ab \rightarrow AB$, where the process starts with one Ab individual.
As discussed previously, this extinction probability within $t-1$ time steps can be recursively calculated by its probability generating function
\begin{align}
  f_{Ab}^{\circ (t-1)} & = d_{Ab} + (1-b_{Ab}-d_{Ab})f_{Ab}^{\circ (t-2)} + b_{Ab}\left((1-\mu_B^A) f_{Ab}^{\circ (t-2)} + \mu_B^A e_{AB}\right)^2,
\end{align}
with $f_{Ab}^{\circ (0)} = 1$.
Similarly, the extinction probability for the subprocess $aB \rightarrow AB$ within $t-2$ time steps can be calculated recursively using the probability generating function for aB
\begin{align}
 f_{aB}^{\circ (t-2)} & = d_{aB} + (1-b_{aB}-d_{aB})f_{aB}^{\circ (t-3)} + b_{aB}\left((1-\mu_A^B) f_{aB}^{\circ (t-3)} + \mu_A^B e_{AB}\right)^2,
\end{align}
with $f_{aB}^{\circ (0)} = 1$.
When we now consider the extinction probability of the whole process starting with an individual of type ab, we see that it can either go extinct right away, or if it divides we can refer to the individual extinction probabilities for the different types (in case of mutation), i.e.\ their probability generating functions
\begin{align}
 \bar{f}_{ab}^{\circ (t)} & := d_{ab} + (1-b_{ab}-d_{ab})\bar{f}_{ab}^{\circ (t-1)} + b_{ab}\left((1-\mu_A-\mu_B)f_{ab}^{\circ (t-1)} + \mu_A f_{Ab}^{\circ (t-1)} + \mu_B f_{aB}^{\circ (t-2)} \right)^2 \nonumber \\
  &  = \bar{f}_{ab} (\bar{f}_{ab}^{\circ (t-1)}, f_{Ab}^{\circ (t-1)}, f_{aB}^{\circ (t-2)}),
\end{align}
with $\bar{f}_{ab}^{\circ (0)} = 1$, $f_{Ab}^{\circ (0)} = 1$, and $f_{aB}^{\circ (0)} = 1$.
Note, that in contrast to the normal probability generating function, here the probability generating function for type aB has one time step less, which agrees with the second term in \ref{eq:probs}.
To not confuse this modified probability generating function with the common one, we use the \textit{bar-notation}.
Again, no probability generating function for the $AB$-type is necessary, since the actual extinction probability for this type is used.

We define this recursive function as
\begin{align}
 \bar{f}_{ab}^{\circ (t)}(s_{ab},s_{Ab}, s_{aB}, s_{AB}) := \bar{f}^{(Ab)} (t).
\end{align}
The index $Ab$ denotes, that this is the modified probability generating function for the pathway via $Ab$.

With this we now describe the algorithm for the path probability.

\begin{enumerate}
  \item Calculate the extinction probability of the final mutant type $AB$ as above.
  
  \item Until some $t_{max}$ calculate recursively $f(t)$ as explained above in Eq. \ref{eq:timedistri}.
  
  \item Until some $t_{max}$ calculate recursively
  
  \begin{align}
    f_{aB}^{\circ (t)} & = d_{aB} + (1-b_{aB}-d_{aB})f_{aB}^{\circ (t-1)} + b_{aB}\left((1-\mu_A^B) f_{aB}^{\circ (t-1)} 
    + \mu_A^B e_{AB}\right)^2, \nonumber \\
    f_{Ab}^{\circ (t)} & = d_{Ab} + (1-b_{Ab}-d_{Ab})f_{Ab}^{\circ (t-1)} + b_{Ab}\left((1-\mu_B^A) f_{Ab}^{\circ (t-1)} 
    + \mu_B^A e_{AB}\right)^2, \\
    \bar{f}^{(Ab)}(t) &:=\bar{f}_{ab}^{\circ (t)} = d_{ab} + (1-b_{ab}-d_{ab})\bar{f}_{ab}^{\circ (t-1)} \nonumber \\
               & + b_{ab}\left((1-\mu_A-\mu_B)\bar{f}_{ab}^{\circ (t-1)} + \mu_A f_{Ab}^{\circ (t-1)} + \mu_B f_{aB}^{\circ (t-\textcolor{red}{2})} \nonumber\right)^2
  \end{align}
  where $f_{aB}^0 = f_{aB}^{-1} = f_{Ab}^0 = f_{ab}^0 = 1$.
  Note, that the only difference is that the probability generating function of types not along the pathway considered is one time step behind (marked in red). This is also the reason, why there are two initial conditions needed for type $aB$.
  
  \item The probability to get the final, successful $AB$ mutant exactly at time $t$ via path $Ab$ and not getting a successful $AB$ mutant beforehand is then computed by
  \begin{align}
    \rho_{Ab} = f^N(t-1) - \left( \bar{f}^{(Ab)}(t)\right) ^N.
  \end{align}
\end{enumerate}

Analogously one can calculate the path probability for reaching the final mutant via $aB$.
Note, that while this computation gives the correct path probabilities, the sum over all paths can be slightly greater than the overall time distribution.
This is due to the fact, that in time discrete systems the final mutant can be reached by different pathways at the same time.
In the description here, such cases count for all pathways that succeed at the time.

\bibliographystyle{naturemag}

\end{document}